\documentstyle[11pt,paspconf,epsf]{article}

\begin{document}

\title{Light Element Settling in Main-Sequence Pop II Stars and the
Primordial Lithium Abundance}
\author{Sylvie Vauclair}
\affil{Laboratoire d'Astrophysique\ Observatoire Midi-Pyr\'en\'ees \
Toulouse, France}

\begin{abstract}

While theoretical computations of the lithium abundances in Pop II stars predict variations
from star to star, the most recent observations confirm the existence of a ``lithium
plateau" with no abundance dispersion. This apparent paradox has to be faced and a solution
has to be found before deriving the primordial lithium abundance. In any case lithium 7 has
been depleted in these stars by at least 20 percent and lithium 6 by at
least a factor 2, even in the hottest stars of the ``plateau". Consequences for other
elements are also discussed. 

\end{abstract}


\section{Introduction}

Since the first observations of the ``lithium plateau" in main-sequence
Pop II stars by Spite and Spite 1982, many abundance determinations have
confirmed the constancy of the lithium abundance in most of these stars.
Moreover the dispersion around the average value is extremely small,
below the observational errors (see Molaro 1999 for an extensive review
on this subject).

On the other hand, the theory of stellar structure and evolution,
including the best available physics, has stongly improved these last
few years. New equations of state, opacities, nuclear reaction rates are
included, as well as the element settling which occurs in radiative
zones. This last process is indeed considered as a ``standard process" as
it represents an improvement in the physics of stellar structure without
any arbitrary parameter. Furthermore, helioseismology gives a
spectacular evidence that stellar physics is improving : the agreement
of the sound velocity in the models and in the ``seismic Sun" (deduced
from helioseismic modes) is much better when element settling is included
(Figure 1).

\begin{figure}
\epsfysize=8cm
\epsfbox{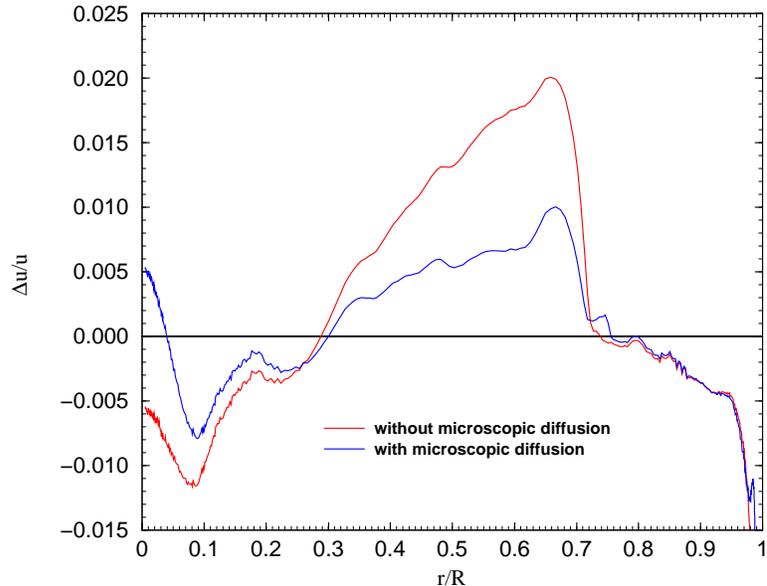}
\caption{Comparison of the $u = P/\rho$ function in the ``seismic Sun"
as obtained from helioseismology by the Warsaw group and from  
our models. Dashed line : without settling; solid line : with element
settling computed for helium and 14 other elements(after Richard et al.,
1999)}
\end{figure}

When applied to main-sequence Pop II stars, these standard models lead
to a lithium depletion increasing with effective temperature, 
in contradiction with the observations.

Two attitudes are then possible :
1) forget about physics and claim that, as lithium is the same in all
these stars, it must be the primordial one.
2) take physics into account ;
in this case we are in the presence of a paradox as lithium
should vary from star to star, which is not observed.

Solving this paradox may lead to important scientific improvements.
What does the lithium plateau want to tell us that we have not yet
understood? We are not ready to answer this question, 
but we can try to find some clues.

\section{The lithium 7 behavior}

The computations of lithium abundance profiles inside standard stellar models lead to a
characteristic behavior which is now wellknown, and which can be seen in Figure 3.
Two reasons merge to lead to lithium 
depletion : gravitational and thermal settling below the convective zone, nuclear
destruction underneath. 

The discrepancy between the predicted lithium values at present time (here with 
an assumed age of 12
Gyr) and the observed value is shown on Figure 2. This paradox suggests that some
non-standard process acts in the stars to prevent the settling effect.

\begin{figure}
\epsfysize=8cm
\epsfbox{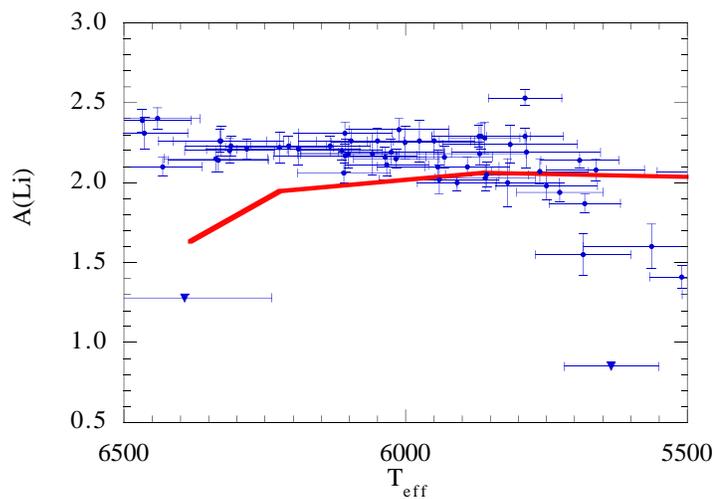}
\caption{Lithium abundances predicted by standard models after 12 Gyr as
a function of effective temperature (after Vauclair and Charbonnel
1995), compared to the observations of Bonifacio and Molaro 1998}
\end{figure}

A first step towards a solution has been given by Vauclair and
Charbonnel 1998 (VC98) who showed that in standard models an ``abundance
attractor" exists for lithium 7 : the maximum lithium abundance which
remains inside the stars has the same value for all the
effective temperatures of the plateau stars, which may be easily
understood in terms of time scales (Figure 3). This maximum occurs when the settling time
scale and the nuclear destruction time scale are the same.

\begin{figure}
\epsfysize=6cm
\epsfbox{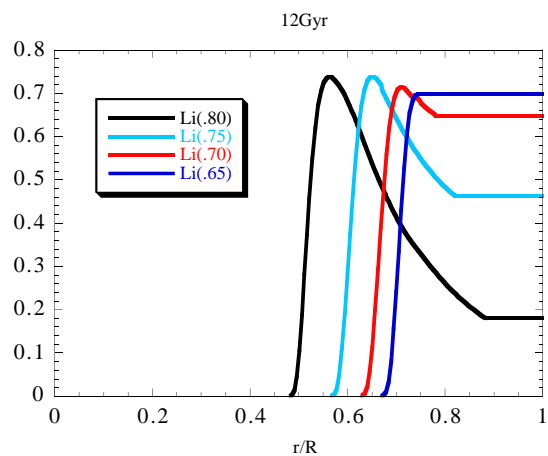}
\caption{ Lithium 7 abundance profiles in main-sequence Pop II stars as
a function of the fractional radius; curves are shown for .85, .8, .75
and $.7 M_o$  at the age of 12 Gyrs}
\end{figure}

The constancy of this maximum value in all the ``plateau stars", whatever the effective
temperature and metallicity, while in all other cases the expected lithium values are so
fluctuent is striking. It leads to the idea that
some macroscopic motions mixes matter
between the bottom of the convective zone and the lithium peak region
(Figure 4). It may be mass loss, as proposed by Vauclair and Charbonnel
1995 (VC95) : in this case all the stars should have suffer an average
mass loss rate of a few $10^{-13} M_o$ per year during their lifetimes. Rotation-
induced mixing or mixing by internal waves can also play a role. The
difficulty with these interpretations in that all the stars should have
had similar histories to account for the low dispersion of the lithium
plateau (cf. Pinsonneault 1999).

\begin{figure}
\epsfysize=12cm
\epsfbox{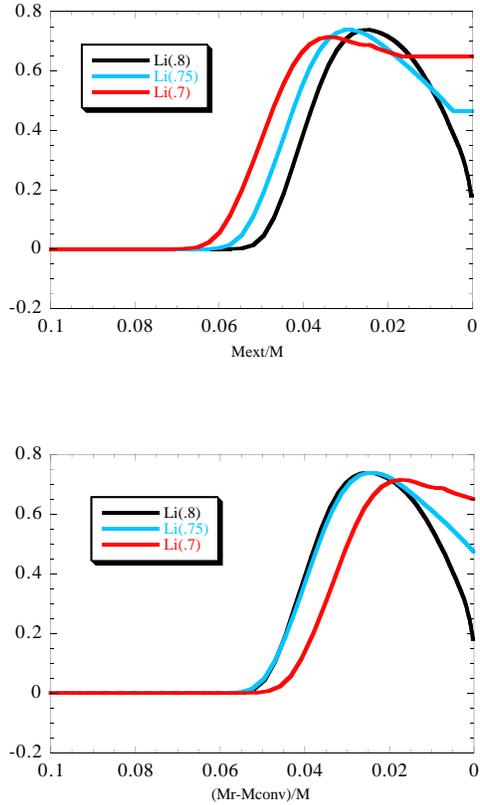}
\caption{ Lithium 7 abundance profiles in main-sequence Pop II stars as
a function of the fractional mass. Left side : curves for .85, .8 and
$.75 M_o$  at the age of 12 Gyrs. Right side : same curves but the
fractional mass of the convective zone is now deduced from the absissa.
We can see that for the ``plateau stars" the lithium peak appears at the
same place below the convection zone. These results suggest that some
mild macroscopic motion mixes matter between the convective zone and the
lithium peak to account for the observations}
\end{figure}

\section{conclusion and consequences for other light elements}

It can be seen from the computations that in any case the 
lithium abundance observed in Pop II stars cannot be the
original one. It has been depleted by at least 20 percent. The smallest
depletion would occur if some
macroscopic motion mixes matter below the convection zone down to the peak value.
 In all other cases the depletion is larger. 
From the
computations of the lithium profiles in standard models and the constancy of the peak value,
we expect that the observed value is indeed related to the peak abundance and that the
depletion is small.

Although we do not yet have any definitive answer on the lithium 7
problem, we may already derive some conclusions for other
elements (Charbonnel and Vauclair 1999). Figure 5 gives the $^6$Li , Be and
B abundance profiles in a $.8 M_o$ stars as well as $^7$Li. We can see that
in any case, if $^7$Li is not depleted by more than 20 percent, the $^6$Li
abundance must have decreased by at least a factor 2. This is an important
result : contrary to current belief, $^6$Li cannot have retained its
original abundance in the hottest stars of the plateau.

\begin{figure}
\epsfysize=6cm
\epsfbox{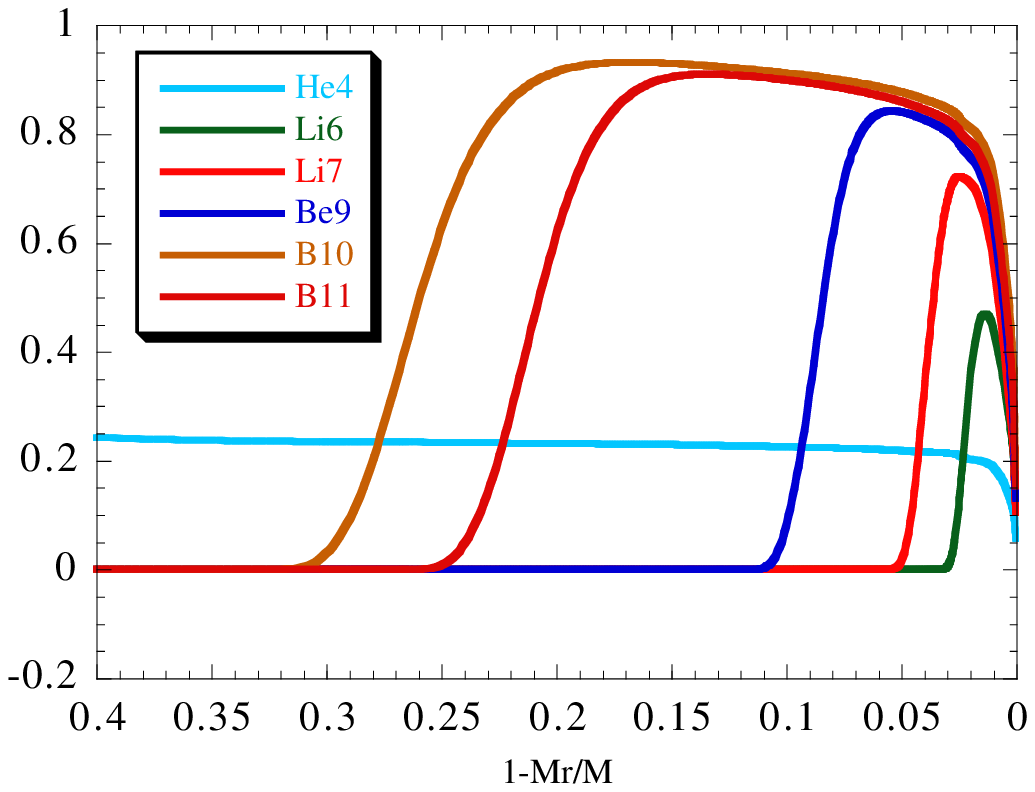}
\caption{ Lithium, beryllium and boron abundance profiles in a $.8 M_o$ main 
sequence 
Pop II stars as
a function of the fractional radius.
We can see that, if $^7$Li is not depleted by more than 20 percent, then $^6$Li 
must be depleted by at
least a factor 2.
Be and B are expected to be depleted by also about 20 percent}
\end{figure}

Similarly, as seen on Figure 5, a depletion of Be and B by about 20
percent is also expected. Should these be less depleted, it would mean
that the macroscopic motions go further down to prevent their depletion,
and then $^7$Li would fall in the region of nuclear destruction.

The few stars which lay far below the plateau (with only upper limits in
lithium) can still be accounted for in two extreme ways : they can
either be the result of specially large macroscopic motions below the
convective zone - the depletion would then be due to nuclear destruction
- or be the result of specially small motions - the depletion would then
be due to settling. Observations of lithium dispersion around the
turn-off in globular clusters like M92 could give a clue in this respect
(Charbonnel and Vauclair 1999). More work is still needed on the light
element evolution in old stars to understand correctly their behavior
and they can, in turn, give us important informations about
hydrodynamical processes.

\end{document}